\documentclass[conference,10pt]{IEEEtran}
\IEEEoverridecommandlockouts
\usepackage{cite}
\usepackage{amsmath,amssymb,amsfonts}
\usepackage{algorithmic}
\usepackage{graphicx,color}
\usepackage{textcomp}
\def\BibTeX{{\rm B\kern-.05em{\sc i\kern-.025em b}\kern-.08em
    T\kern-.1667em\lower.7ex\hbox{E}\kern-.125emX}}

\usepackage{anyfontsize}
\usepackage{tabulary}
\usepackage[table]{xcolor}
\usepackage{multirow,multicol}
\usepackage{array,ragged2e}
\usepackage{tikz,soul}
\usepackage[nodisplayskipstretch]{setspace}
\usepackage{mathtools} 
\usepackage{xspace}

\newcommand{\subtext}[1]{\text{\fontfamily{cmr}\fontshape{n}\fontseries{m}\selectfont{}#1}}
\newcommand{\sub}[1]{\ensuremath{_{\subtext{#1}}}} 

\newcommand{\Figref}[1]{Fig.~\ref{#1}}

\newcommand{\bh}{\mathbf{h}}

\newcommand{\bq}{\mathbf{q}}

\newcommand{\bw}{\mathbf{w}}
\newcommand{\by}{\mathbf{y}}
\newcommand{\bx}{\mathbf{x}}

\newcommand{\bC}{\mathbf{C}}
\newcommand{\bD}{\mathbf{D}}

\newcommand{\bI}{\mathbf{I}}

\newcommand{\bR}{\mathbf{R}}

\newcommand{\bzero}{\boldsymbol{0}}

\newcommand{\herm}{^{\mathsf{H}}}
\newcommand{\trans}{^\mathsf{T}}

\DeclareMathOperator{\tr}{tr}
\DeclareMathOperator{\E}{\mathbb{E}}

\DeclareMathOperator{\Pro}{\mathsf{Pr}}

\newcommand{\EX}[1]{\E\left\{{#1}\right\}}

\newcommand{\Prx}[1]{\Pro\left\{{#1}\right\}}

\DeclareMathOperator*{\argmin}{arg\,min}

\newcommand{\norm}[1]{{ \left\Vert #1 \right\Vert }}

\newcommand{\mC}{\mathbb{C}}

\newcommand{\Bc}{B_{\mathrm{c}}}
\newcommand{\Tc}{T_{\mathrm{c}}}

\newcommand{\nd}{n_{\mathrm{d}}}
\newcommand{\nvu}{\sigma^2_{\mathrm{u}}}
\newcommand{\nvd}{\sigma^2_{\mathrm{d}}}

\newcommand{\iid}{\text{i.i.d.}}
\newcommand{\CN}[2]{\mathcal{CN}\left({#1},{#2}\right)}

\newcommand{\Atj}{\tilde{A}^t_j}

\newcommand{\rhoe}{\rho^{\mathsf{e}}}
\newcommand{\rhou}{\rho^{\mathsf{u}}}

\newcommand{\rhoMax}{\rho^{\mathsf{max}}}

\newcommand{\setKe}{\mathcal{K}^{\mathrm{e}}}
\newcommand{\setKu}{\mathcal{K}^{\mathrm{u}}}
\newcommand{\au}{a_{\mathrm{u}}}
\newcommand{\datae}{\varsigma^{\mathsf{e}}}
\newcommand{\datau}{\varsigma^{\mathsf{u}}}
\newcommand{\datas}{\varsigma^{\mathsf{s}}}

\makeatletter
\def\@setsize#1#2#3#4{
    \@nomath#1
    \let\@currsize#1
    \baselineskip #2
    \baselineskip \baselinestretch\baselineskip
    \parskip \baselinestretch\parskip
    \setbox\strutbox \hbox{
        \vrule height.7\baselineskip
            depth.3\baselineskip
            width\z@}
    \skip\footins \baselinestretch\skip\footins
    \normalbaselineskip\baselineskip#3#4}
\makeatother


\DeclareRobustCommand{\IEEEauthorrefmark}[1]{\smash{\textsuperscript{\footnotesize #1}}}

\begin{document}

\title{On the Coexistence of eMBB and URLLC in the Cell-Free Massive MIMO Downlink\\
\thanks{This work was supported by the European Union under the Italian National Recovery and Resilience Plan (NRRP) of NextGenerationEU, partnership on ``Telecommunications of the Future'' (PE00000001 - program ``RESTART'', Structural Project 6GWINET, Cascade Call SPARKS, CUP D43C22003080001).}
}

\author{\IEEEauthorblockN{Giovanni Interdonato\IEEEauthorrefmark{1,2}, Stefano Buzzi\IEEEauthorrefmark{1,2,3}}
\IEEEauthorblockA{\IEEEauthorrefmark{1}\textit{Department of Electrical and Information Engineering, University of Cassino and Southern Latium, 03043 Cassino, Italy.}\\
\IEEEauthorrefmark{2}\textit{Consorzio Nazionale Interuniversitario per le Telecomunicazioni (CNIT), 43124 Parma, Italy.}\\
\IEEEauthorrefmark{3}\textit{Politecnico di Milano, 20133 Milan, Italy.}\\
\texttt{\{giovanni.interdonato,buzzi\}@unicas.it}}
}


\newcounter{counter_1}

\begin{figure*}[t!]
\normalsize
This paper has been submitted for publication in the proceedings of an IEEE conference.

\

\textcopyright~2024 IEEE. Personal use of this material is permitted. 
Permission from IEEE must be obtained for all other uses, in any current or future media, including reprinting/republishing this material for advertising or promotional purposes, creating new collective works, for resale or redistribution to servers or lists, or reuse of any copyrighted component of this work in other works.

\


\vspace{17cm}
\end{figure*}
 
\maketitle

\begin{abstract}
We investigate the non-orthogonal coexistence between the ultra-reliable low-latency communication (URLLC) and the enhanced mobile broadband (eMBB) in the downlink of a cell-free massive multiple-input multiple-output (MIMO) system. 
We provide a unified information-theoretic framework that combines a finite-blocklength analysis of the URLLC error probability based on the use of mismatched decoding with an infinite-blocklength analysis of the eMBB spectral efficiency. Superposition coding and three levels of puncturing are considered as alternative downlink coexistence strategies to cope with the inter-service interference and the URLLC random activation pattern, under the assumption of imperfect pilot-based channel state information acquisition at the access points and statistical channel knowledge at the users. 
Numerical results shed light into the trade-off between eMBB and URLLC performances considering different precoding and power control strategies.
\end{abstract}


\section{Introduction}

Academic research and industrial standardization are currently exploring different coexistence mechanisms between \textit{enhanced mobile broadband} (eMBB) and \textit{ultra-reliable low-latency communications} (URLLCs) aiming at simultaneously meeting their ``conflicting'' requirements, and apparently diverging from the initial vision of an \textit{orthogonally-sliced} radio access network (RAN).    
Heterogeneous non-orthogonal multiple access (H-NOMA) can be attained by letting heterogeneous services share the same time-frequency resources, while separating them in the power and/or spatial domain.

A first tractable communication-theoretic model that captures the key features of the coexistence between eMBB and URLLC was proposed in~\cite{Popovski2018slicing}.
In~\cite{Esswie2018} a spatial preemptive scheduler for joint URLLC and eMBB traffic is proposed for multi-user multiple-input multiple-output (MIMO) systems. A similar study but for the DL of a distributed fog network was conducted in~\cite{Abedin2019}. \cite{Kassab2019} provides an information-theoretic study in the performance of URLLC and eMBB, under H-NOMA, in cloud-RAN assuming fading channels and the lack of channel state information (CSI) at the URLLC transmitters.
%
However, the information-theoretic framework used by the aforementioned works hinges on the use of the outage capacity to characterize the performance of the URLLC and leads to an inaccurate and overly optimistic evaluation of the error probability~\cite{Ostman2021}. 
As an alternative, finite-blocklength analyses have been proposed for URLLC in conventional cellular networks~\cite{Saggese2022}, co-located massive MIMO networks~\cite{Ren2020} and cell-free massive MIMO (CF-mMIMO) networks~\cite{Nasir2021}, relying on the information-theoretic bounds developed in~\cite{Polyanskiy2010}, known as the \textit{normal approximation}.
However, the normal approximation is not accurate  in the region of low error probabilities of interest in URLLC (i.e., $<\!10^{-4}$)~\cite{Ostman2021}, especially  in presence of imperfect CSI. The work in~\cite{Ostman2021} provided a more rigorous finite-blocklength information-theoretic framework relying on the use of a mismatched decoding, and of the \textit{saddlepoint approximation} for evaluating the error probability of URLLCs in co-located massive MIMO systems. While the analysis of~\cite{Ostman2021} is limited to the URLLC regime, in~\cite{Interdonato2023} the coexistence between eMBB and URLLC has been investigated under a unified information-theoretic framework that combines the infinite-blocklength analysis for the eMBB ergodic rate and the finite-blocklength analysis for the URLLC error probability. The works in~\cite{Lancho2023, Huang2024} extends the framework introduced in~\cite{Ostman2021} to the case of CF-mMIMO, but limit their analysis to the URLLC regime only. Recent attempts of studying the eMBB-URLLC coexistence in CF-mMIMO systems still rely either on the outage capacity~\cite{Bucci2024} or on the rate via the normal approximation~\cite{Liu2023}.
Therefore, a proper unified information-theoretic framework that sheds light into the design of CF-mMIMO architectures able to support the eMBB-URLLC coexistence has still to be defined.

\textbf{Contributions:} We extend the information-theoretic framework introduced in~\cite{Interdonato2023} to investigate the coexistence between eMBB and URLLC in CF-mMIMO systems.
Unlike prior works~\cite{Lancho2023,Huang2024,Bucci2024,Liu2023} wherein the URLLC users are coherently served by multiple access points (APs), we realistically assume that, due to the strict latency constraints, the URLLC traffic is locally generated at the single AP serving the URLLC user. The proposed framework accommodates four alternative coexistence strategies: \textit{superposition coding} (SPC), local, cluster-based and network-wide \textit{puncturing} (PUNC). 
The analysis accounts for imperfect pilot-based CSI acquisition, URLLC random activation pattern, spatially correlated fading and the lack of CSI at the users. 
We evaluate the performance achieved by PUNC and SPC under maximum-ratio and minimum mean square error precoding with weighted fractional power control.    


\section{System Model}
\label{sec:system-model}

Let us consider a CF-mMIMO system operating in time-division duplexing (TDD) mode and at sub-6 GHz frequency bands. $L$ APs, equipped with $M$ antennas each, are distributed over a square area of $D\!\times\!D$ km$^2$ and connected through a fronthaul network to a central edge processor (CEP). The APs coherently serve $K$ single-antenna users (UEs) in the same time-frequency resource, with $LM\!\gg\!K$.
 
The conventional block-fading channel model is considered. Let $\tau_c$ denote the channel coherence block length, and $\tau_c \!=\! \Tc \Bc$ samples (or \textit{channel uses}) where $\Tc$ is the coherence time and $\Bc$ is the coherence bandwidth. 
We assume a coherence block accommodates UL training and DL data transmission. 
Hence, $\tau_c\!=\!\tau_p\!+\!\tau_d$, where $\tau_p$ and $\tau_d$ are the training and the DL data transmission duration, respectively.
The coherence time is, in turn, divided in $T$ slots of equal duration. A slot consists of $\nd$ channel uses, with $\nd \!=\! \lfloor \tau_d/T \rfloor$.
%
The set including the indices of the eMBB and URLLC UEs is denoted as $\setKe$ and $\setKu$, respectively.
The eMBB traffic is generated at the CEP, enabling an arbitrary eMBB UE $k$ to be coherently served by a subset of the APs in the network denoted by $\mathcal{L}_k \subseteq \{1,\ldots,L\}$. While, the set $\mathcal{U}_l$ denotes the set of the UE indices served by AP $l$. An eMBB transmission spans multiple coherence blocks, wherein the channel realizations evolve independently according to the block-fading model.\footnote{Theoretically, an eMBB transmission spans an infinite number of coherence block according to the ergodic regime.}
Conversely, due to the low latency requirements, the URLLC traffic is generated locally at the AP serving the URLLC UE of interest. Thus, there is no coherent cell-free transmission towards URLLC UEs, but rather a small-cell-like approach wherein a URLLC UE is served by its \textit{master} AP only. Each URLLC transmission follows a random activation pattern and is confined within a single slot to meet the very strict latency requirements. Hence, the number of channel uses in a slot equals the URLLC codeword length. Within a coherence block, a URLLC user may be active in multiple slots.
The AP-to-UE service association is mathematically handled with a set of diagonal matrices $\bD_{kl} \in \mC^{M \times M}$, for $k=1,\ldots,K$ and $l=1,\ldots,L$, where $\bD_{kl}$ is the identity matrix ${\bf I}_M$ if AP $l$ participates in the service of UE $k$, and ${\bf 0}_{M\times M}$ otherwise. 

The channel between the $k$-th UE and the $l$-th AP is denoted by $\bh_{kl}\!\in\! \mC^{M}$, with $\bh_{kl}\!\sim\!\CN{\bzero}{\bR_{kl}}$, and $\bR_{kl}\! \in\! \mC^{M \times M}$ being the spatial correlation matrix. The corresponding large-scale fading coefficient is defined as $\beta_{kl}\! =\! \tr(\bR_{kl})/M$ and captures the path loss and shadowing effects. 
In the UL training phase, each AP estimates the channels of the UEs it is supposed to serve according to the cooperation clusters established by the matrices $\{\bD_{kl}\}$. Due to space limitations, we omit details about the UL training phase and refer to~\cite[Sec. 4.2]{cellfreebook}. We let $\widehat{\bh}_{kl}$ be the MMSE estimate of $\bh_{kl}$.
The estimation error $\widetilde{\bh}_{kl} = \bh_{kl} \!-\! \widehat{\bh}_{kl}$ is uncorrelated to the estimate and distributed as $\widetilde{\bh}_{kl} \!\sim\!\CN{\bzero_N}{\bC_{kl}}$, with $\bC_{kl}$ being the error correlation matrix.
We reasonably assume that correlation matrices are known wherever required.

\subsection{eMBB-URLLC Coexistence Strategies}
To handle the eMBB-URLLC coexistence, we propose the following DL transmission techniques: $(i)$ local puncturing (LPu), $(ii)$ cluster-based puncturing (CPu), $(iii)$ network-wide puncturing (NPu), and $(iv)$ superposition coding (SPC).
Under LPu, an AP that triggers a URLLC (dubbed \textit{critical} AP) in a certain slot punctures all its eMBB transmissions therein. The eMBB service can be still guaranteed to the punctured eMBB UEs by other serving APs. 
Under CPu, whenever a URLLC is triggered by a critical AP in a certain slot, all the eMBB transmissions from all the APs belonging to the same user-centric cluster of the URLLC UE are dropped. The eMBB service can be still guaranteed to the punctured UEs by other out-of-cluster serving APs.    
Under NPu, whenever a URLLC is triggered by a critical AP in a certain slot, all the eMBB transmissions throughout the cell-free network are dropped. 
The eMBB service may be guaranteed to the punctured UEs in the remaining slots where no URLLC UEs are active. 
Under superposition coding, eMBB transmissions occur in all the slots and each critical AP linearly combines eMBB and URLLC signals whenever URLLCs are triggered.
 
Let $A^t_{kj}$ be a binary coefficient that equals 1 if a URLLC transmission takes place at the $t$-th slot between AP $j$ and URLLC UE $k$, and 0 otherwise. This coefficient models the random activation pattern of the URLLC users which follows a Bernoulli distribution with parameter $\au$, $A^t_{kj} \!\sim\!\mathcal{B}\mathit{ern}(\au)$.
For an arbitrary AP $j$ and slot $t$, the aforementioned techniques can be mathematically handled by the coefficient
\begin{align}
\Atj = \begin{cases}
{\bigg(1-\sum\limits_{\ell \in \mathcal{L}} \sum\limits_{i \in \setKu_{\ell}} A^t_{i\ell} \bigg)}^{\!\!+}, \quad& \text{for PUNC}, \\[-.2em]
1 \ , \quad& \text{for SPC}, 
\end{cases}
\end{align}
where $\setKu_{\ell} \!=\! \setKu \cap \mathcal{U}_{\ell}$ and $x^+\!=\!\max\{x,0\}$. Besides, $\mathcal{L} \!=\! \{j\}$ for LPu, $\mathcal{L} \!=\!\mathcal{L}_{\mathcal{U}_j}$ for CPu, and $\mathcal{L} \!=\! \{1,\ldots,L\}$ for NPu. The notation $\mathcal{L}_{\mathcal{U}_l}\subseteq \{1,\ldots,L\}$ denotes the set of the (unique) indices of the APs serving the UEs that are served by AP $j$ ($\mathcal{L}_{\mathcal{U}_l}$ includes AP $j$).   
Let $\datae_{k}[n]$ or $\datau_{k}[n]$ be the data symbol intended for UE $k$ over the $n$-th channel use, if $k$ is an eMBB UE or a URLLC UE, respectively.
We assume that $\datas_{k}[n] \!\sim\! \CN{0}{1}$, with $\mathsf{s} = \{\mathsf{e},\mathsf{u}\}$.    
AP $j$ precodes the data symbol by using the precoding vector $\bw_{kj}\!\in\!\mC^{M}$, $\E\{\norm{\bw_{kj}}^2\}\!=\!1$, which is function of the acquired CSI. The data signal transmitted by AP $j$ over an arbitrary channel use $n$ of slot $t$~is
\begin{align}
\bx^t_j[n] &\!=\! \Atj \sum\nolimits_{k \in \setKe} \sqrt{\rhoe_{kj}} \bD_{kj}\bw_{kj}\datae_{k}[n] \nonumber \\
 &\!\quad+ \sum\nolimits_{i \in \setKu} A^t_{ij} \sqrt{\rhou_{ij}} \bD_{ij}\bw_{ij}\datau_{i}[n],   
\end{align}
with $n=1,\ldots,\nd$, and where $\rhoe_{kj}$ and $\rhou_{ij}$ are the DL transmit powers used by AP $j$ to its eMBB UE $k$ and URLLC UE $i$, respectively, satisfying the following per-AP power constraint
\begin{align}
\EX{\norm{\bx^t_j[n]}^2} \!=\! \Atj \sum\limits_{k \in \setKe_j}  {\rhoe_{kj}} \!+\! \sum\limits_{i \in \setKu_j}  A^t_{ij} {\rhou_{ij}} \leq \rhoMax_j,    
\label{eq:data-power-constraint}
\end{align}
with $j \!=\! 1,\ldots,L$, $\setKe_j \!=\! \setKe \cap \mathcal{U}_j$, and where $\rhoMax_j$ is the maximum transmit power at AP $j$.
The data signal received at UE $k$ over an arbitrary channel use $n$ of slot $t$ is denoted as $y^{t,\mathsf{s}}_{k}[n]$, with $\mathsf{s} = \{\mathsf{e},\mathsf{u}\}$. 
If UE $k \in \setKe$, then its signal received over channel use $n$ of slot $t$ can be written as in~\eqref{eq:received-data-signal-eMBB} at the top of the next page, where $w_{k}[n]\!\sim\!\CN{0}{\nvd}$ is the \iid~receiver noise with variance $\nvd$, and we have defined $\varrho^t_{ij} \!=\! A^t_{ij} \sqrt{\rhou_{ij}}$, if $i \!\in\! \setKu$, or $\varrho^t_{ij} \!=\! \Atj \sqrt{\rhoe_{ij}}$, if $i \!\in\! \setKe$. Besides, $g_{kil} \!=\! \bh\herm_{kl}\bD_{il}\bw_{il}$, namely the effective DL channel between AP $l$ and user $k$, precoded by the precoding vector intended for user $i$. 
%
\begin{figure*}[!t]
\normalsize
\setcounter{counter_1}{\value{equation}}
\setcounter{equation}{8}
\begin{align}
\label{eq:received-data-signal-eMBB}
y^{t,{\mathsf{e}}}_{k}[n] &\!=\! \underbrace{\vphantom{\sum\limits_{i \in \setKe\setminus\{k\}}}\datae_{k}[n]\sum^L_{j=1}\varrho^t_{kj}\EX{g_{kkj}}}_{\text{desired signal}} + \underbrace{\vphantom{\sum\limits_{i \in \setKe\setminus\{k\}}} \datae_{k}[n]\sum^L_{j=1} \varrho^t_{kj} ({g_{kkj}} \!-\! \EX{g_{kkj}})}_{\text{self-interference}} +\!\! \underbrace{\sum\limits_{i \in \setKe\setminus\{k\}} \!\!\!\datae_{i}[n] \sum^L_{j=1} \varrho^t_{ij} g_{kij}}_{\text{intra-service interference}} \!+\!\! \underbrace{\sum\limits_{i \in \setKu} \!\datau_{i}[n] \sum^L_{j=1} \varrho^t_{ij} g_{kij}}_{\text{inter-service interference}} \!+\! \underbrace{\vphantom{\sum\limits_{i \in \setKe\setminus\{k\}}}w_{k}[n]}_{\text{noise}}\,,
\end{align}
\setcounter{equation}{\value{equation}}
\hrulefill
\vspace*{-6mm}
\end{figure*}
%
If UE $k$ is a URLLC UE, its signal received over channel use $n$ in slot $t$ can be written as in~\eqref{eq:received-data-signal-URLLC} at the top of the next page. 
Eq.~\eqref{eq:received-data-signal-eMBB} emphasizes the fact that UE $k$  solely knows the statistical CSI of the DL channel, that is $\EX{g_{kkj}}$. The second term in~\eqref{eq:received-data-signal-eMBB} represents the self-interference due to this lack of CSI, referred to as \textit{beamforming gain uncertainty}. 

\begin{figure*}[!t]
\normalsize
\setcounter{counter_1}{\value{equation}}
\setcounter{equation}{9}
\begin{align}
\label{eq:received-data-signal-URLLC}
y^{t,{\mathsf{u}}}_{k}[n] \!=\! \underbrace{\vphantom{\sum\limits_{i \in \setKu\setminus\{k\}}} \datau_{k}[n] \sum^L_{j=1} \varrho^t_{kj} g_{kkj}}_{\text{desired signal}} 
+\!\! \underbrace{\sum\limits_{i \in \setKu\setminus\{k\}} \!\! \datau_{i}[n] \sum^L_{j=1} \varrho^t_{ij} g_{kij} }_{\text{intra-service interference}}
+\!\underbrace{\vphantom{\sum\limits_{i \in \setKu\setminus\{k\}}} \sum\limits_{i \in \setKe} \datae_{i}[n] \sum\limits^L_{j=1} \varrho^t_{ij} g_{kij}}_{\text{inter-service interference}}
+ \underbrace{\vphantom{\sum\limits_{i \in \setKu\setminus\{k\}}}w_{k}[n]}_{\text{noise}}\,,
\end{align}  
\setcounter{equation}{\value{equation}}
\hrulefill
\vspace*{-3mm}
\end{figure*}

\subsection{Precoding and Power Allocation}
\label{subsec:precoding-power-control}
We realistically assume that each AP designs the precoding vectors on a coherence block basis rather than on a slot basis. Besides, the precoding scheme design is agnostic with respect to the random activation pattern of the URLLC UEs and the coexistence strategy. The precoders are designed by assuming the worst case in which all the URLLC UEs are active in all the slots. 
We herein consider two precoding schemes.
\textit{Local Partial MMSE (LP-MMSE)}~\cite{cellfreebook} suppresses only the interference caused to the UEs in the set $\mathcal{U}_j$. It consists in setting $\bw^{\mathsf{LP-MMSE}}_{kj} \!=\! \bar{\bw}_{kj}/\sqrt{\E\{\norm{\bar{\bw}_{kj}}^2\}}$ with
\begin{equation*}
\bar{\bw}_{kj} \!=\!  p_{k}  \left( \sum\nolimits_{i \in \mathcal{U}_j} p_{i} \big( \widehat{\bh}_{ij} \widehat{\bh}_{ij}\herm \!+\! \bC_{ij} \big) \!+\! \nvu  \bI_{N} \right)^{\!\!-1} \!\!\bD_{kj} \widehat{\bh}_{kj},
\end{equation*}
where $\{p_i\}_{i=1}^K$ are the UE transmit powers. 
\textit{(Normalized) Maximum-Ratio (MR)} is computationally the cheapest, but performance-wise the worst, precoding scheme and consists in setting $\bw_{kj}^{\mathsf{MR}} \!=\! \widehat{\bh}_{kj}/||\widehat{\bh}_{kj}||.$ 
%
Power allocation can be heuristically carried out on a slot basis by taking into account the adopted eMBB-URLLC coexistence strategy and the URLLC activation pattern, which is known at the AP in the DL operation. In this paper, we re-adapt the local \textit{weighted fractional power allocation} (FPA) introduced in~\cite{Interdonato2023} by setting 
\begin{align}
\rhou_{ij} &\!=\!  \frac{\omega \rhoMax_j A^t_{ij} {(\beta_{ij})}^{\nu}}{(1\!-\!\omega)\Atj \!\sum\limits_{k \in \setKe_j}\! {(\beta_{kj})}^{\nu} \!\!+\! \omega\!\!\sum\limits_{u \in \setKu_j} \!\! A^t_{uj}{(\beta_{uj})}^{\nu}},\,i\!\in\!\setKu_j\,,  \\[-.5ex]
\rhoe_{kj} &\!=\! \frac{(1-\omega) \rhoMax_j \Atj {(\beta_{kj})}^{\nu}}{(1\!-\!\omega)\Atj \sum\limits_{e \in \setKe_j}\! {(\beta_{ej})}^{\nu} \!\!+\! \omega\!\! \sum\limits_{i \in \setKu_j} \!\!A^t_{ij} {(\beta_{ij})}^{\nu}},\,k\!\in\!\setKe_j\,,
\end{align}
where the weight $\omega \in (0,1)$ tunes the amount of DL power allocated to the URLLC UEs, while $\nu$ determines the policy as a function of the average channel gain. An opportunistic power allocation is attained by setting $\nu\!>\!0$, with which more power is allocated to the UEs with better channel conditions. Conversely, fairness is supported by setting $\nu\!<\!0$. If $\omega\in (0.5,1)$ a larger share of power is allocated to the URLLC UEs than to the eMBB UEs. If $\nu\!=\!0$ and $\omega\!=\!0.5$, then the FPA reduces to an equal power allocation (EPA).
Notably, FPA satisfies the per-AP power constraint in~\eqref{eq:data-power-constraint} with equality.

\section{Performance Analysis}
\label{sec:performance-analysis}
We evaluate the spectral efficiency (SE) achieved by the eMBB UEs in a single coherence block by hinging on to the information-theoretic bounds and tools in the infinite-blocklength regime~\cite{cellfreebook}.
An achievable DL SE can be obtained by applying the popular \textit{hardening bound} technique~\cite{cellfreebook} on the signal model in~\eqref{eq:received-data-signal-eMBB}, by treating all the interference sources as uncorrelated noise. Specifically, an achievable DL SE of an arbitrary eMBB UE $k$, is given by 
\begin{align}
\label{eq:eMBB:SE}
\mathsf{SE}^{\mathsf{e}}_{k} = \dfrac{\tau_d}{\tau_c} \dfrac{1}{T}\sum^T_{t=1} \log_2(1+\gamma^{t,\mathsf{e}}_{k}), \text{ [bits/s/Hz]} \ ,
\end{align}
where $\tau_d/\tau_c$ is the fraction of channel uses spent on the DL transmission, and the effective SINR of UE $k \in \setKe$ is
{\color{black} 
\begin{align} \label{eq:eMBB:SINR}
\!\!\gamma^{t,\mathsf{e}}_{k} \!=\! \frac{\left|\sum\nolimits^L_{j=1}\varrho^t_{kj}\EX{g_{kkj}}\right|^2}{\!\sum\limits^{K}_{i=1}\!\EX{\left|\sum\limits^L_{j=1} \varrho^t_{ij} g_{kij}\right|^2}\!-\!\left|\sum\limits^L_{j=1}\varrho^t_{kj}\EX{g_{kkj}}\right|^2\!\!\!+\!\nvd},
\end{align}
}%
where the expectations are taken with respect to the random channel realizations.
The expression of the achievable SE shown in~\eqref{eq:eMBB:SE} holds for any choice of coexistence technique between heterogeneous services, other than for any choice of precoding scheme and any channel distribution.

The error probability of the URLLC transmissions is characterized on a slot basis by applying the finite-blocklength information-theoretic results established in~\cite{Lancho2023}.
Let us define 
\begin{align}
\widetilde{g}_{ki} \!=\! \sum\nolimits^L_{j=1} \varrho^t_{ij} \bh\herm_{kj}\bD_{ij}\bw_{ij} = \bh\herm_{k}\widetilde{\bD}^{t}_{i}\bw_{i}\, , 
\end{align} 
where $\bh_k \!=\! [\bh\trans_{k1}~\cdots~\bh\trans_{kL}]\trans$, $\bw_k \!=\! [\bw_{k1}~\cdots~\bw_{kL}]\trans$, and $\widetilde{\bD}^{t}_i \!=\! \text{blkdiag}( \widetilde{\bD}^{t}_{ij} , \ldots, \widetilde{\bD}^{t}_{iL} )$, with $\widetilde{\bD}^{t}_{ij}\!=\!A^t_{ij}\sqrt{\rhou_{ij}} \bI_M$ if $j \!\in\! \mathcal{L}_k$, and $\bzero_{M \times M}$ otherwise. 
Hence,~\eqref{eq:received-data-signal-URLLC} can be rewritten as $y^{t,{\mathsf{u}}}_{k}[n] \!=\! \widetilde{g}_{kk} q_{k}[n]  \!+\! z_{k}[n], \, n=1,\dots,\nd,$
where, for the sake of brevity, $q_{k}[n] \!=\! \datau_{k}[n]$, and 
\begin{align}
\!z_{k}[n] \!= \!\!\sum\limits_{i \in \setKu\setminus\{k\}} \!\!\widetilde{g}_{ki} q_{i}[n] \!+\!\! \sum\limits_{i \in \setKe} \!\datae_{i}[n] \sum\limits^L_{j=1} \varrho^t_{ij} g_{kij} \!+\! w_{k}[n] \, . 
\end{align} 
However, URLLC UE $k$ has not access to $\widetilde{g}_{kk}$, but rather to its mean value, $\widehat{g}_{kk} \!=\! \E\{\bh\herm_{k}\widetilde{\bD}^{t}_{k}\bw_{k}\}$, which is treated as perfect. 
Notice that, $\widetilde{g}_{kk}$ remains constant for any other transmission to UE $k$ over slots in the same coherence block.  
Given all channels and precoding vectors, the effective noise terms $\{z_{k}[n]\!\in\!\mathbb{C}; n=1,\ldots,\nd\}$ are $\CN{0}{\sigma_{k}^2}$, with variance
\begin{align}
\sigma_{k}^2 &= \sum\limits_{i \in \setKu\setminus\{k\}} |\widetilde{g}_{ki}|^2 \!+ \!\sum\limits_{i \in \setKe} \Bigg|\sum\limits^L_{j=1} \Atj \sqrt{\rhoe_{ij}} g_{kij}\Bigg|^2  + \nvd \ . 
\end{align}
To decode the transmitted codeword $\bq_{k}\!=\![q_{k}[1],\ldots,q_{k}[\nd]]\trans $, UE $k$ employs a \textit{mismatched scaled nearest-neighbor} (SNN) decoder, with which selects the codeword $\widetilde{\bq}_{k}$ from the codebook $\mathcal {C}$ by applying the rule 
\begin{equation}\label{eq:mismatched_snn_decoder}
\widehat{\bq}_{k}=\argmin_{\widetilde{\bq}_{k}\in\mathcal{C}} \norm{\by^{t,{\mathsf{u}}}_{k}-\widehat{g}_{kk}\widetilde{\bq}_{k}}^2 \ ,
\end{equation}
where $\by^{t,{\mathsf{u}}}_{k} \!=\! [y^{t,{\mathsf{u}}}_{k}[1],\dots,y^{t,{\mathsf{u}}}_{k}[\nd]]\trans$ is the received data vector.
Let $\epsilon^{\mathsf{dl}}_{k} \!=\! \Prx{\widehat{\bq}_{k}\neq \bq_{k}}$ be the DL error probability experienced by the URLLC UE $k$. An upper bound on $\epsilon^{\mathsf{dl}}_{k}$ is  obtained by using the standard \textit{random-coding} approach~\cite{Lancho2023},
\begin{align}
\label{eq:rcus_fading}
\!\epsilon^{\mathsf{dl}}_{k} \!\leq\! \EX{\!\Prx{\sum_{n=1}^{\nd} {\imath_s(q_{k}[n],\by^{t,{\mathsf{u}}}_{k}[n])} \!\leq\! \log\frac{m\!-\!1}{r} \bigg| g_{kk}}\!\!}, 
\end{align}
where $m \!=\! 2^b$ is the number of codewords with length $\nd$ that convey $b$ information bits, $r$ is a random variable uniformly distributed in the interval $[0,1]$ and $\imath_s(q_{k}[n],\by^{t,{\mathsf{u}}}_{k}[n])$ is the \textit{generalized information density}. 
In~\eqref{eq:rcus_fading} the expectation is taken over the distribution of $g_{kk}$, and the probability is computed with respect to the DL data symbol $\{q_{k}[n]\}_{n=1}^{\nd}$, the effective additive noise $\{z_{k}[n]\}_{n=1}^{\nd}$, and the random variable $r$.
The upper bound in~\eqref{eq:rcus_fading} can be reliably evaluated by using the \textit{saddlepoint} approximation provided in~\cite[Sec. II-C]{Lancho2023}.\footnote{The saddlepoint approximation is more accurate than the normal approximation in the URLLC massive MIMO regime~\cite{Ostman2021}. It characterizes the exponential decay of the error probability, as a function of the URLLC codeword length, and the transmission rate requirement $R\!=\!(\log m)/\nd$. Details are omitted due to the lack of space.}
The \textit{DL network availability} can be obtained as $\eta^{\mathsf{dl}} \!=\! \mathsf{Pr}\{\epsilon^{\mathsf{dl}}_{k} \leq \epsilon^{\mathsf{dl}}\sub{trg}\}$, where the probability is taken with respect to the large-scale fading, and measures the probability that the target error probability, $\epsilon^{\mathsf{dl}}\sub{trg}$, is satisfied by an arbitrary UE $k$.

\section{Simulation Results}
\label{sec:simulation-results}
In our simulations, we consider a CF-mMIMO system with $L\!=\!100$ APs, each one equipped with a uniform linear array with equispaced half-wavelength antenna elements. The users are dropped uniformly at random over the coverage area with $D\!=\!1$ km. 
The channel correlation matrices $\{\bR_{kl}\}$ are generated according to the \textit{local scattering} spatial correlation model~\cite[Sec. 2.5.3]{cellfreebook}, assuming jointly Gaussian angular distributions of the multipath components around the nominal azimuth and elevation angles. The random variations in the azimuth and elevation angles are assumed to be independent, and the corresponding angular standard deviations are equal to 15$^{\circ}$ (i.e., strong spatial channel correlation). The average channel gain is obtained according to the 3GPP Urban Microcell model defined in \cite[Table B.1.2.1-1]{LTE2017}. 
We consider $B\!=\!20$ MHz, receiver noise power equal to -94 dBm, $\rhoMax\!=\! 200$ mW, the UL transmit powers equal to 100 mW for all the UEs, the URLLC packet consists of $b\!=\!160$ bits and $\epsilon^{\mathsf{dl}}\sub{trg}\!=\!10^{-5}$.
We set $\tau_c = 580$ channel uses, given by $\Tc\!=\!2$ ms and $\Bc\!=\!290$ kHz.\footnote{With these settings, we mainly target URLLC applications such as motion control for factory automation and closed-loop control for process automation that require a minimum latency of 2 ms~\cite{URLLCspecs3GPP}.} Lastly, $\tau_p \!=\! 10$, and pilots are re-used among the UEs. To assign the pilot as sparsely as possible in space, we adopt the joint pilot assignment and clustering in~\cite[Sec. 5.4]{cellfreebook}. The master AP for each UE is selected as the AP with the largest average channel gain.   
\begin{figure}[!t]
\centerline{\includegraphics[width=.75\columnwidth]{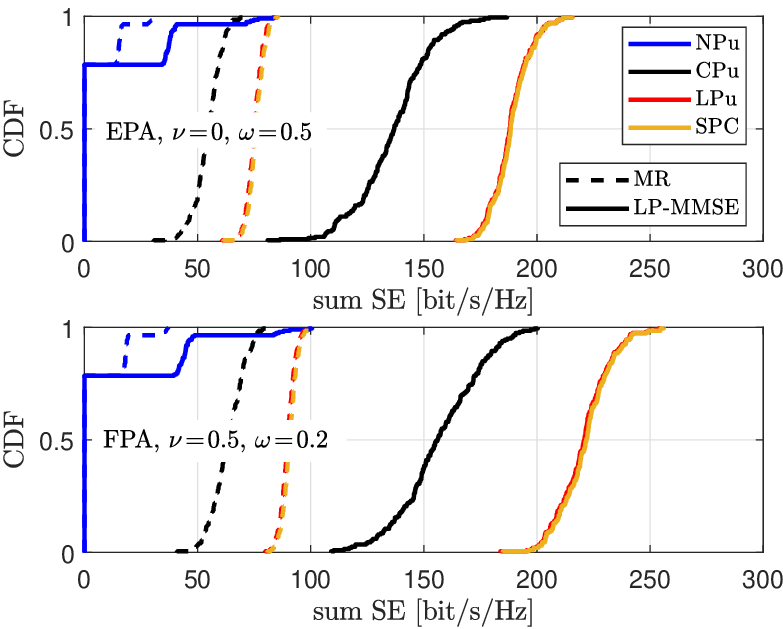}}
\vspace*{-3mm}
\caption{CDFs of the DL sum SE achieved by LP-MMSE and MR with FPA ($\nu\!=\!0.5$, $\omega\!=\!0.2$) and EPA. $K\!=\!40$, $\alpha\!=\!0.2$, $\au\!=\!10^{-0.5}$, $T\!=\!5$, $\nd\!=\!114$.}
\label{fig:fig1}
\vspace*{-2mm}
\end{figure}
In these simulations we consider: $K\!=\!40$, $\alpha\!=\!0.2$, $\au\!=\!10^{-0.5}$, $T\!=\!5$ slots of length $\nd\!=\!114$ channel uses. 
In~\Figref{fig:fig1} we plot the cumulative distribution function (CDF) of the DL sum SE achieved by LP-MMSE and MR with FPA ($\nu\!=\!0.5$, $\omega\!=\!0.2$) and EPA ($\nu\!=\!0$, $\omega\!=\!0.5$), under SPC and the considered three levels of PUNC (LPu, CPu and NPu).        
As expected, SPC and LPu are greatly superior than CPu and NPu. Interestingly, the SE degradation introduced by LPu is negligible. LP-MMSE provides up to $2.5\times$ higher SE than MR, while FPA can bring an additional SE gain up to 18\% compared to EPA by allocating 30\% more power to the eMBB UEs with an opportunistic allocation strategy. Notably, these results point out the eMBB \textit{service outage} that likely occurs when NPu is adopted, namely $\varsigma_{\, \mathrm{out}} \!=\! \Prx{\sum_{k \in \setKe} \mathsf{SE}^{\mathsf{e}}_{k} \!=\! 0}$. This probability of providing no service in a coherence block to eMBB UEs depends on the URLLC activation pattern, the number of URLLC UEs and the number of slots per coherence block. Under the settings considered in~\Figref{fig:fig1}, the eMBB service outage is quite significant as amounts to about 80\%. 

\begin{figure}[!t]
\centerline{\includegraphics[width=.75\columnwidth]{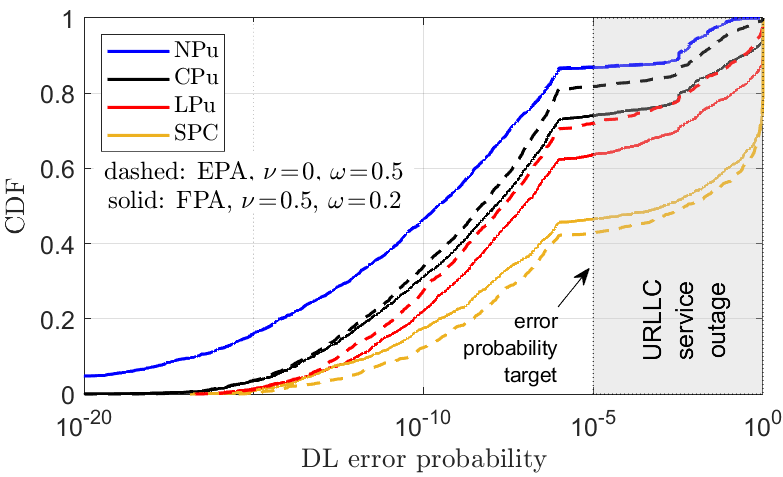}}
\vspace*{-3mm}
\caption{DL per-user error probability achieved by LP-MMSE. Settings are those reported in ~\Figref{fig:fig1}. The error probability target is $\epsilon^{\mathsf{dl}}\sub{trg}\!=\!10^{-5}$.}
\label{fig:fig2}
\vspace*{-5mm}
\end{figure}

In~\Figref{fig:fig2} we show the CDFs of the DL error probability achieved by LP-MMSE with FPA and EPA.
The network availability is identified by the cross-point between the CDF of the DL error probability and the vertical line representing the error probability target value.
The interference caused by the eMBB UEs to the URLLC UEs, when SPC is performed, is significant and leads to a network availability value below 0.5. CPu guarantees slightly higher error probability than CPu, while NPu provides the best URLLC performance. Notably, EPA outperforms FPA concerning PUNC as 30\% more power is allocated with EPA to the URLLC UEs than to the eMBB UEs. Lastly, in this scenario MR provides URLLC service outage, namely network availability equal to zero, regardless of the coexistence strategy (hence results are omitted). 
From these simulations, we conclude that LPu ensures the best trade-off between eMBB SE and URLLC network availability, and that, in such a critical interference-limited scenario, LP-MMSE is vital to prevent a large URLLC service outage.

To observe a significant improvement in the network availability, especially for MR, we should consider $M\!=\!16$ antennas per AP, a FPA with $\nu\!=\!0.5$, $\omega\!=\!0.8$, and $T\!=\!2$ slots. The latter leads to $\nd\!=\!285$, thereby reducing the URLLC rate requirement, $R\!=\!b/\nd$.%
\begin{figure}[!t]
\centerline{\includegraphics[width=.75\columnwidth]{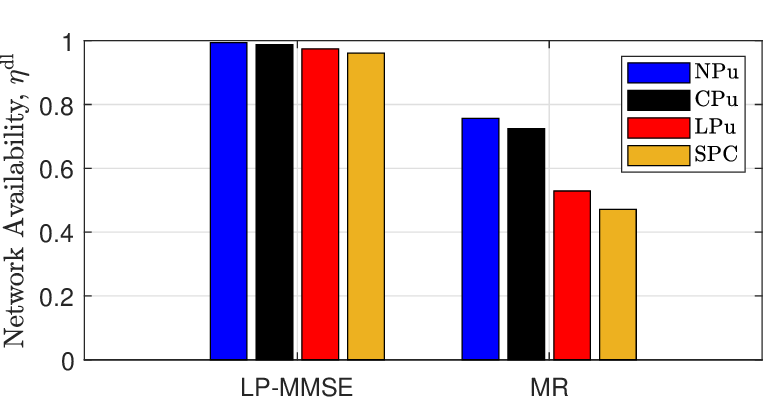}}
\vspace*{-3mm}
\caption{Network availability achieved by LP-MMSE and MR with FPA, $\nu\!=\!0.5$, $\omega\!=\!0.8$. Here, $K\!=\!40$, $M\!=\!16$, $\alpha\!=\!0.2$, $T\!=\!2$, $\nd\!=\!285$.}
\label{fig:fig3}
\vspace*{-5mm}
\end{figure}
In this scenario, we observe from~\Figref{fig:fig3} that any of the proposed coexistence strategies is able to provide levels of network availability close to 1, provided that LP-MMSE is used. Indeed, increasing the number of AP antennas entails an enhanced ability to cancel the intra-AP interference. On the other hand, the network availability achieved by MR lies in the interval 0.45--0.8, thanks to the increased array gain and an increased power allocated to the URLLC UEs. Either way, the decrease of the DL error probability is also due to a looser URLLC rate requirement set by the slot length, $\nd$.

\section{Conclusion}
\label{sec:conclusion}
We investigated the coexistence between eMBB and URLLC in the DL of a CF-mMIMO system by presenting an information-theoretic framework that combines both asymptotic and nonasymptotic analyses to evaluate the performance of eMBB and URLLC, respectively.
This framework accommodates alternative strategies: superposition coding (SPC), local, cluster-based and network-wide puncturing (PUNC).

Simulation results revealed that the spatial degrees of freedom available at the APs are crucial to resolve the interference caused by SPC. However, PUNC is necessary to keep the URLLC error probability below the strict target, at the expense of some eMBB spectral efficiency (SE) loss. In this regard, local PUNC ensures the best trade-off between eMBB SE and URLLC network availability.
Extensions to this work may concern the study of coexistence strategies for the
uplink, also including massive machine-type communications.



\bibliographystyle{IEEEtran}
\bibliography{IEEEabrv,refs-abbr}

\end{document}